\documentclass[10pt]{article}

 \usepackage{amsmath,amsthm,amssymb}
 \usepackage{makeidx,epsfig,lscape}
 \usepackage{color,colortbl}
 \usepackage{fancyhdr}
 \usepackage{xcolor,pict2e}

 \definecolor{myaqua}{rgb}{0.0,0.5,0.55}
 \definecolor{lightaqua}{rgb}{0.75,0.95,0.95}

 \usepackage[colorlinks = true,
            linkcolor = myaqua,
            urlcolor  = blue,
            citecolor = myaqua]{hyperref}

\usepackage{caption}
\usepackage{floatrow}
\newcommand{\ket}[1]{\mbox{$|#1\protect\rangle$}}
\newcommand{\bra}[1]{\mbox{$\protect\langle#1|$}}

\def\lin#1#2{\textcolor[rgb]{0.6,0.6,0.6}{\vspace*{#1mm} \hrule
   height 3 pt \vspace*{#2mm}}}
\def\bt{\begin{tabular}}
\def\et{\end{tabular}}
\def\and{\mbox{ and }}

\def\1{{\bf 1}}

 \def\sectionn#1{\refstepcounter{section}{\color{myaqua}

 \vskip 6mm

 \noindent\Large\bf\thesection. #1}

 \vskip 3mm}

\begin{document}

 \fancyhead[L]{\hspace*{-13mm}
 \bt{l}{\bf Journal of
Quantum Information Science, 2015, *,**}\\
 Published Online **** 2014 in SciRes.
 \href{http://www.scirp.org/journal/*****}{\color{blue}{\underline{\smash{http://www.scirp.org/journal/****}}}} \\
 \href{http://dx.doi.org/10.4236/****.2014.*****}{\color{blue}{\underline{\smash{http://dx.doi.org/10.4236/****.2014.*****}}}} \\
 \et}
 \fancyhead[R]{\includegraphics{pic1.ps}}

 $\mbox{ }$

 \vskip 12mm

{ 

{\noindent{\huge\bf\color{myaqua}
  A Holevo-type bound for a Hilbert Schmidt distance measure}}
%
\\[6mm]
{\large\bf Boaz Tamir$^1$, Eliahu Cohen$^2$}}
\\[2mm]
{ 
 $^1$Faculty of interdisciplinary studies, Bar-Ilan University, Ramat-Gan, Israel\\
Email: \href{mailto:canjlm@actcom.co.il}{\color{blue}{\underline{\smash{canjlm@actcom.co.il}}}}\\[1mm]
$^2$School of Physics and Astronomy, Tel-Aviv University, Tel-Aviv, Israel\\
Email:
\href{mailto:eliahuco@post.tau.ac.il}{\color{blue}{\underline{\smash{eliahuco@post.tau.ac.il}}}},
 \\[4mm]
Received **/**/****
 \\[4mm]
Copyright \copyright \ 2015 by authors and Scientific Research Publishing Inc. \\
This work is licensed under the Creative Commons Attribution International License (CC BY). \\
\href{http://creativecommons.org/licenses/by/4.0/}{\color{blue}{\underline{\smash{http://creativecommons.org/licenses/by/4.0/}}}}\\
 \includegraphics{pic2.ps}

\lin{5}{7}

 { 
 {\noindent{\large\bf\color{myaqua} Abstract}{\bf \\[3mm]
 \textup{We prove a new version of the Holevo bound employing the Hilbert-Schmidt norm instead of the Kullback-Leibler divergence. Suppose Alice is sending classical information to Bob using a quantum channel, while Bob is performing some projective measurement. We bound the classical mutual information in terms of the Hilbert-Schmidt norm by its quantum Hilbert-Schmidt counterpart. This constitutes a Holevo-type upper bound on the classical information transmission rate via a quantum channel. The resulting inequality is rather natural and intuitive relating classical and quantum expressions using the same measure.
 }}}
 \\[4mm]
 {\noindent{\large\bf\color{myaqua} Keywords}{\bf \\[3mm]
 Holevo bound; Hilbert-Schmidt norm; entanglement measures
}

 \fancyfoot[L]{{\noindent{\color{myaqua}{\bf How to cite this
 paper:}} B. Tamir and E. Cohen (2015)
 A Holevo-type bound for a Hilbert Schmidt distance measure.
 {\it Journal of Quantum Information Science},*,***-***}}

\lin{3}{1}

\sectionn{Introduction and Motivation}

{ \fontfamily{times}\selectfont
 \noindent Holevo's theorem \cite{Holevo} is one of the pillars of quantum information theory. It can be informally summarized as follows: `It is not possible to communicate more than $n$ classical bits of information by the transmission
of $n$ qubits alone'. It therefore sets a useful upper bound on the classical information rate using quantum channel.

Suppose Alice prepares a state $\rho_x$ in some system $Q$, where $x \in X=\{0,...,n\}$ with probabilities $p(x)$. Bob performs a measurement described by the POVM elements ${E_Y}={E_0,...,E_m}$ on that state, with measurement outcome $Y$. Let

\begin{equation}
\rho =\rho^Q = \sum_x p(x) \rho_x.
\end{equation}

\noindent The Holevo Bound states that \cite{Nielsen}

\begin{equation}
H(X:Y) \le S(\rho)-\sum_x p(x)S(\rho_x),
\end{equation}

\noindent where $S$ is the von Neumann entropy and $H(X:Y)$ is the Shannon mutual information of $X$ and $Y$. Recent proofs of the Holevo bound can be found in \cite{HBP1,HBP2}.  \\

Consider the following trace distance between two probability distributions $p(x)$ and $q(x)$ on $X$ \cite{Remark}

\begin{equation}
d(p||q) = \frac{1}{2} \sum_x (p(x)-q(x))^2.
\end{equation}
\noindent We can extend the definition to density matrices $\rho$ and $\sigma$
\begin{equation}
d(\rho||\sigma) = \frac{1}{2} tr (\rho -\sigma)^2,
\end{equation}

\noindent where we use $A^2$ for $A^\dag A$. This is known as the Hilbert-Schmidt (HS) norm \cite{Buzek,Vedral,Garcia} (in fact this is one half of the HS norm). Recently, the above distance measure was coined the `logical divergence' of two densities \cite{Ellerman}.

We prove a Holevo-type upper bound on the mutual information of $X$ and $Y$, where the mutual information is written this time in terms of the HS norm instead of the Kullback-Leibler divergence. It was recently suggested by Ellerman \cite{Ellerman} that employing the HS norm in the formulation of classical mutual information is natural. This is consistent with the identification of information as a measure of distinction \cite{Ellerman}. Note that employing the Kullback-Leibler divergence in the standard form of the Holevo bound gives an expression which can be identified with quantum mutual information, however the `coherent information' is considered a more appropriate expression (see also \cite{Nielsen} chapter 12).  In view of the above we hereby take Ellerman's idea a step further and write a Holevo-type bound based on the HS norm.

The question whether $d(\rho||\sigma)$ is the right measure of quantum mutual information was discussed in \cite{Wolf}, within the context of area laws. It was used there to provide an upper bound on the correlations between two distant operators $M_A$ and $M_B$, where $A$ is a region inside a spin grid and $B$ is its complement:

\begin{equation}
d(\rho^{A,B}||\rho^A\otimes \rho^B)\geq  \frac{C(M_A,M_B)^2}{2 ||M_A||^2 M_B||^2},
\end{equation}

\noindent where $C(M_A,M_B)=\langle M_A \otimes M_B \rangle - \langle M_A \rangle \langle M_B \rangle $ is the correlation function of $M_A$ and $M_B$.

In addition, the HS norm was suggested as an entanglement measure \cite{Vedral,HS2}, however this was criticized in \cite{HS3}, claiming it does not fulfill the so called CP non-expansive property (i.e. non-increasing under every completely-positive trace-preserving map).

\noindent In what follows we will prove a Holevo-type bound on the above HS distance between the probability $p(x,y)$ on the product space $(X,Y)$ and the product of its marginal probabilities $p(x)\cdot p(y)$:

\begin{equation}  \label{ineq}
\frac{1}{q} d(X,Y||X \otimes Y) \leq  d(\rho^{P,Q}||\rho^P\otimes \rho^Q),
\end{equation}

\noindent \noindent where
\begin{equation}
\rho^{P,Q} = \sum_x p(x) \ket{x}\bra{x}\otimes \rho_x,
\end{equation}

\noindent and where

\begin{equation}
\rho^Q = \sum_x p(x) \rho_x,
\end{equation}

\begin{equation}
\rho^P = \sum_x p(x) \ket{x} \bra{x}
\end{equation}

\noindent  are the partial traces of $\rho^{P,Q}$, and $q$ is the dimension of the space $Q$. Note that both sides of inequality \ref{ineq} are measures of mutual information. Therefore, our claim is that the classical HS mutual information is bounded by the corresponding quantum one multiplied by the dimension of the quantum density matrices used in the channel. We will also show that

\begin{equation}
\frac{1}{q} d(X,Y||X \otimes Y) \leq \frac{1}{2} L(\rho^P) L(\rho^Q),
\end{equation}

\noindent where $L(\rho^Q)$ and $L(\rho^P)$ are the Tsallis entropies \cite{Tsallis}
\begin{equation}
L(\rho) = tr(\rho(1-\rho)),
\end{equation}

\noindent also known as the linear entropy, purity \cite{entropy} or logical entropy of $\rho$ \cite{Ellerman}.


All the above is proved for the case of projective measurements. However we expect similar results in the general case of POVM, in light of Naimark's dilation theorem (see \cite{Pau} or \cite{Hol} for instance).

In the next section we review some basic properties of quantum logical divergence and then use these properties to demonstrate the new Holevo-type bound.

\sectionn{The HS norm and the Holevo-type bound}

{ \fontfamily{times}\selectfont
 \noindent
Let

\begin{equation}
d(\rho||\sigma) = \frac{1}{2} tr (\rho -\sigma)^2.
\end{equation}

In what follows we recall some basic properties of the HS distance measure, then we state and prove the main result of this paper.

{\bf Theorem 2.1: Contractivity of the HS norm with respect to projective measurements:}

Let $\rho$ and $\sigma$ be two density matrices of a system $S$. Let $\mathcal{E} (\rho)$ be the trace preserving operator
\begin{equation}
\mathcal{E} (\rho) = \sum_i P_i\rho P_i ,
\end{equation}

\noindent where the projections $P_i$ satisfy $\sum_iP_i =I$, $P_i^\dag = P_i$ and $P_i^2 = P_i$ for every $i$, then

\begin{equation}
d(\mathcal{E}(\rho)||\mathcal{E}(\sigma))\leq d(\rho||\sigma).
\end{equation}

{\bf Proof:} We now write $X=\rho-\sigma$. Then $X$ is Hermitian with bounded spectrum, and using Lemma 2 in \cite{Rastegin} we conclude that

\begin{equation}
tr(\mathcal{E}\rho -\mathcal{E}\sigma)^2 < tr(\rho-\sigma)^2.
\end{equation}

{\bf Theorem 2.2: The joint convexity of the HS norm}

The logical divergence $d(\rho||\sigma)$ is jointly convex.

{\bf Proof:} First observe that $tr(\rho^2)$ is convex from the convexity of $x^2$ and the linearity of the trace. 
Next we can write
\begin{equation}
d(\lambda \rho_1 + (1-\lambda) \rho_2|| \lambda \sigma_1 + (1-\lambda) \sigma_2)=
\end{equation}
\[=\frac{1}{2}tr((\lambda \rho_1 +(1-\lambda)\rho_2) -(\lambda\sigma_1 + (1-\lambda) \sigma_2))^2 =\]
\[=\frac{1}{2}tr(\lambda(\rho_1-\sigma_1) + (1-\lambda)(\rho_2-\sigma_2))^2 \leq \]
\[\leq \frac{\lambda}{2} tr(\rho_1-\sigma_1)^2 + \frac{1-\lambda}{2} tr(\rho_2-\sigma_2)^2 =\]
\[= \lambda d(\rho_1||\sigma_1) +(1-\lambda) d(\rho_2||\sigma_2), \]

\noindent where the inequality is due to the convexity of $tr(\rho^2)$. This constitute the joint convexity. \\

{\bf Theorem 2.3: The monotonicity of the HS norm with respect to partial trace}

Let $\rho^{A,B}$ and $\sigma^{A,B}$ be two density matrices, then

\begin{equation}
d(\rho^A\otimes I/b||\sigma^A\otimes I/b) \leq d(\rho^{A,B}||\sigma^{A,B}),
\end{equation}
\noindent where $b$ is the dimension of $B$.

{\bf Proof:} One can find a set of unitary matrices $U_j$ over $B$ and a probability distribution $p_j$ such that
\begin{equation}
\rho^A \otimes I/b = \sum_j p_j U_j \rho^{A,B} U_j^\dag
\end{equation}
\begin{equation}
\sigma^A \otimes I/b = \sum_j p_j U_j \sigma^{A,B} U_j^\dag,
\end{equation}
\noindent (see \cite{Nielsen} chapter 11). Now since $d(\rho||\sigma)$ is jointly convex on both densities, we can write

\begin{equation}
d(\rho^A\otimes I/b||\sigma^A\otimes I/b) \leq \sum_{j} p_j  \cdot d(U_j \rho^{A,B} U_j^\dag||U_j \sigma^{A,B} U_j^\dag).
\end{equation}

Observe now that the divergence is invariant under unitary conjugation, and therefore the sum in the right hand side of the above inequality is $d(\rho^{A,B}||\sigma^{A,B})$. \\

We can now state the main result:

{\bf Theorem 2.4: A Holevo-type bound for the HS trace distance between $p(x,y)$ and $p(x) \cdot p(y)$}

Suppose Alice is using a distribution $p(x)$, where $x$ is in  $X= \{ 1,...,n\}$, to pick one of $n$ densities $\rho_x$ in $Q$. She then sends the signal in a quantum physical channel to Bob. We can add an artificial quantum system $P$ and write $(P,Q)$ for Alice as:
\begin{equation}
\label{PQ}
\rho^{P,Q} = \sum p(x) \ket{x}\bra{x}\otimes \rho_x,
\end{equation}
\noindent where the vectors $\ket{x}$ are orthogonal. Let $\rho^P$ and $\rho^Q$ be the partial traces of $\rho^{P,Q}$. Suppose Bob is measuring the system using a projective measurement as in Theorem 2.1, then
\begin{equation}
\frac{1}{q} d(X,Y||X \otimes Y) \leq d(\rho^{P,Q}||\rho^P \otimes \rho^Q),
\end{equation}
\noindent where $q$ is the dimension of the space $Q$.

{\bf Proof:} First we consider one more auxiliary quantum system, namely $M$ for the measurement outcome for Bob. Initially the system $M$ is in the state $M_0 = \ket{0} \bra{0}$. Let $\mathcal{E}$ be the operator defined by Bob's measurement as in Theorem 2.1 above: let  $\{P_y\}_y$ on $Q$ be defined such that $\sum_y P_y =I$ and

\begin{equation}
\mathcal{E} (\rho_x) = \sum_y P_y \rho_x P_y.
\end{equation}

\noindent One can easily extend $\mathcal{E}$ to the space $(Q,M)$ by
\begin{equation}
\mathcal{E} (\rho \otimes \ket{0}\bra{0}) = \sum_y P_y \rho P_y \otimes \ket{y}\bra{y}.
\end{equation}

\noindent This can be done by choosing a set of operators, conjugating $\ket{0}\bra{0})$ to $\ket{y}\bra{y})$. It amounts to writing the measurement result in the space $M$ (see also Ch. 12.1.1 in \cite{Nielsen}). If we now trace out $Q$ we find
\begin{equation}
tr_Q \mathcal{E} (\rho \otimes \ket{0}\bra{0})= \sum_y p(y)\ket{y}\bra{y}.
\end{equation}
\noindent Moreover, $\mathcal{E}$ can be extended to $\rho^{(P,Q,M)}$ by
\begin{equation}
\mathcal{E} \left ( \sum_x p(x) \ket{x}\bra{x}\otimes \rho_x \otimes \ket{0}\bra{0}\right ) =
\end{equation}
\[ =\sum_x \sum_y p(x) \ket{x}\bra{x}\otimes P_y \rho_x P_y \otimes \ket{y}\bra{y}.\]
\noindent If we trace out $Q$ we arrive at
\begin{equation}
\label{a}
tr_Q \mathcal{E} (\rho^{(P,Q,M_0)}) =
\end{equation}
\[ = \sum_{x,y} p(y/x) p(x) \ket{x}\bra{x} \otimes  \ket{y}\bra{y}\]

\[ = \sum_{x,y} p(x,y) \ket{x}\bra{x} \otimes  \ket{y}\bra{y}. \]

\noindent Finally, we can extend $\mathcal{E}$ to $\rho^P \otimes \rho^{(Q,M)}$ by
\begin{equation}
\mathcal{E} \left ( \sum_x p(x) \ket{x}\bra{x}\otimes \rho \otimes \ket{0}\bra{0}\right ) =
\end{equation}
\[ =\sum_x \sum_y  p(x) \ket{x}\bra{x}\otimes P_y \rho P_y \otimes \ket{y}\bra{y}.\]

\noindent If we trace out $Q$ we get
\begin{equation}
\label{b}
tr_Q \mathcal{E} (\rho^P \otimes \rho^{(Q,M_0)})=
\end{equation}
\[ = \sum_x \sum_y  p_x \ket{x}\bra{x}\otimes p(y) \ket{y}\bra{y}=\]

\[=(\sum_x  p_x \ket{x}\bra{x}) \otimes(\sum_y p_y \ket{y}\bra{y}). \]

We can now use the properties stated in the above theorems and Eqs. $\ref{a}$ and $\ref{b}$ to deduce
\begin{equation}
d(\rho^{P,Q}||\rho^{P}\otimes \rho^{Q})=d(\rho^{P,Q,M_0}||\rho^{P}\otimes \rho^{Q,M_0})\geq
\end{equation}
\[ \geq d(\mathcal{E} (\rho^{P,Q,M_0})||\mathcal{E} (\rho^{P}\otimes \rho^{Q,M_0})) \geq \]
\[ \geq d(tr_Q \mathcal{E} (\rho^{P,Q,M_0})\otimes I_q|| tr_Q \mathcal{E}( \rho^P \otimes \rho^{(Q,M_0)}) \otimes I_q) =\]
\[ = \frac{1}{q} d(X,Y||X \otimes Y), \]

\noindent where in the first inequality we have used Theorem 2.1 and in the second inequality Theorem 2.3. The final equality is an easy consequence of the definition of the HS norm. \\

{\bf Corollary:} Suppose Alice is sending classical information to Bob using a quantum channel $Q$, Bob measures the quantum state using a projective measurement defined above (having results in space $Y$).
Under all the above assumptions
\begin{equation}
\frac{1}{q} d(X,Y||X \otimes Y) \leq \frac{1}{2} L(\rho^P) L(\rho^Q),
\end{equation}
where $L(\rho^Q)$ and $L(\rho^P)$ are Tsallis entropies of the second type (the quantum logical entropies) of $\rho^Q $ and $\rho^P $.\\

{\bf Proof:} Clearly (see also \cite{Tamir})
\begin{equation}
d(\rho^{P,Q}|| \rho^P \otimes \rho^Q) = tr( \rho^{P,Q}(1-\rho^P \otimes\rho^Q))-
\end{equation}
\[ - \frac{1}{2} L(\rho^P \otimes \rho^Q)-\frac{1}{2} L(\rho^{P,Q}).  \]

\noindent It is easy to see (by a matrix representation) that for $\rho^{P,Q}$ as in Eq. $\ref{PQ}$
\begin{equation}
tr( \rho^{P,Q}(1-\rho^P \otimes\rho^Q))= L(\rho^{P,Q}),
\end{equation}
\noindent therefore
\begin{equation}
d(\rho^{P,Q}|| \rho^P \otimes \rho^Q) = \frac{1}{2} L(\rho^{P,Q}) - \frac{1}{2} L(\rho^P \otimes\rho^Q).
\end{equation}

\noindent However, $L(\rho^{P,Q}) \leq L(\rho^P) +L(\rho^Q)$, and $L(\rho^P \otimes\rho^Q)= L(\rho^P) + L(\rho^Q)-L(\rho^P)\cdot L(\rho^Q)$, (see \cite{Tamir} Theorem II.2.4 and Theorem II.3), therefore
\begin{equation}
d(\rho^{P,Q}|| \rho^P \otimes \rho^Q)\leq \frac{1}{2}L(\rho^P)\cdot L(\rho^Q).
\end{equation}

\noindent Combining this with Theorem 2.4 we find

\begin{equation}
\frac{1}{q} d(X,Y||X \otimes Y) \leq \frac{1}{2} L(\rho^P) L(\rho^Q).
\end{equation}

{\bf Example:} Suppose Alice sends the state $\ket{0}$ with probability $1/2$ and the state $\ket{\psi} = \cos \theta \ket{0} + \sin \theta \ket{1}$ with probability $1/2$, then

\begin{equation}
\rho^{P,Q}= \frac{1}{2} \ket{0} \bra{0}\otimes \rho_0 + \frac{1}{2} \ket{1} \bra{1} \otimes \rho_1,
\end{equation}

\noindent where $\rho_0 = \ket{0}\bra{0}$ and $\rho_1 = \ket{\psi}\bra{\psi}$. By partial tracing we get

\begin{equation}
\rho^Q = \frac{1}{2} \left( \begin{array}{cc} 1&0\\0&0 \end{array} \right) +  \frac{1}{2} \left( \begin{array}{cc} \cos^2 \theta & \cos \theta \sin \theta \\ cos \theta \sin \theta & \sin^2 \theta \end{array} \right),
\end{equation}
\noindent and  $\rho^P$ is a balanced coin. The eigenvalues of $\rho^Q$ are $(1 \pm \cos \theta)/2$ and therefore
\begin{equation}
L (\rho^Q) = 1- [(1+ \cos \theta)^2/4 + (1- \cos \theta)^2/4] = \frac{1}{2}\sin^2 \theta.
\end{equation}
\noindent Also $L (\rho^P)= 1/2 $ and $q=2$, hence
\begin{equation}
d(X,Y||X \otimes Y) \leq \frac{1}{4}\sin^2 \theta \leq 1/4.
\end{equation}

The left hand side of the above inequality is a measure of the classical mutual information according to the HS norm between $X$ and $Y$. The very fact that it is smaller than the Tsallis information measure of $X$ (which is $1/2$) means that the quantum channel restricts the rate of classical information transfer, where the mutual information is measured by the HS norm and the source of information $X$ is measured by Tsallis entropy. This is analogous to Holevo's upper bound in the framework of Tsalis/linear entropy. We find this result similar in spirit to the well-known limitation on the rate of classical information transmission via a quantum channel (without utilizing entanglement): one cannot send more than one bit for each use of the channel using a one qubit channel.

In the above example, if $\rho_0$ and $\rho_1$ are mixed states, then by the same argument we can show that:

\begin{equation}
d(X,Y||X\otimes Y) \leq \frac{1}{2} L(\rho^Q)
\end{equation}

\noindent  This gives a bound on the classical mutual information using the quantum `logical entropy' (the Tsallis entropy).

\sectionn{Discussion}
\label{sec:MLE}


{ \fontfamily{times}\selectfont
 \noindent
We proved a Holevo-type bound employing the Hilbert-Schmidt distance between the density matrices on the product space $(X,Y)$ and the tensor of the two marginal density matrices on $X \otimes Y$. Using a different  measure of mutual information, we showed that this Holevo-type upper bound on classical information transmission can be written as an inequality between the classical mutual information expression and its quantum counterpart.

It seems that by utilizing Naimark's dilation \cite{Pau,Hol}, the above result can be generalized to any POVM, if one is willing to employ the suitable channel in a higher dimensional Hilbert space.

As was claimed in \cite{Tamir}, the divergence distance used above is the natural one in the context of quantum logical entropy \cite{Ellerman}. Being the `right' measure of mutual information in quantum channels passing classical information, we expect that this formalism would be helpful in further studying various problems such as channel capacity theory, entanglement detection and area laws.

 {\color{myaqua}

 \vskip 6mm

 \noindent\Large\bf Acknowledgments}

 \vskip 3mm

{ \fontfamily{times}\selectfont
 \noindent
E.C. was partially supported by Israel Science Foundation Grant No. 1311/14.

 {\color{myaqua}

}}

\end{document}